\documentstyle{mn}
\input{epsf}

\def\B{{\rm B}}

\def\df{{\rm d}}
\def\ev{{\rm eV}}
\def\ex{{\rm e}}
\def\H{{\rm H}}
\def\I{{\rm I}}
\def\rec{{\rm rec}}

\begin{document}
\journal{UUITP-1-98, OUTP-98-24-P, SISSA 37/98/A, astro-ph/9805108}
\title[CMB Anisotropy in Decaying Neutrino Cosmology]
      {CMB Anisotropy in the Decaying Neutrino Cosmology}
\author[J.A. Adams, S. Sarkar and D.W. Sciama] 
       {J. A. Adams$^1$, S. Sarkar$^2$ and D. W. Sciama$^3$\\
       $^1$Department of Theoretical Physics, Uppsala University,
              Box 803, S-75108, Uppsala, Sweden\\
       $^2$Theoretical Physics, University of Oxford, 1 Keble Road,
              Oxford OX1 3NP, UK\\
       $^3$S.I.S.S.A., I.C.T.P., Strada Costiera 11, 34014 Trieste, Italy 
              \& Nuclear and Astrophysics Laboratory, University of Oxford}
\date{\today}
\pubyear{1998}
\maketitle
\begin{abstract}
It is attractive to suppose for several astrophysical reasons that the
universe has close to the critical density in light ($\sim30$~eV)
neutrinos which decay radiatively with a lifetime of
$\sim10^{23}$~sec. In such a cosmology the universe is reionized early
and the last scattering surface of the cosmic microwave background
significantly broadened. We calculate the resulting angular power
spectrum of temperature fluctuations in the cosmic microwave
background. As expected the acoustic peaks are significantly damped
relative to the standard case. This would allow a definitive test of
the decaying neutrino cosmology with the forthcoming MAP and PLANCK
surveyor missions.
\end{abstract}
\begin{keywords}
{cosmology: dark matter, reionization, microwave background}
\end{keywords}

\section{Introduction}

The absence of Gunn-Peterson absorption by diffuse neutral hydrogen in
quasar spectra \cite{gp65} indicates that the intergalactic medium
(IGM) is highly ionized, out to a redshift of at least 5
\cite{giallongo}. The integrated flux of ionizing UV photons due to
quasars themselves is believed to be inadequate for this purpose
\cite{gs96,m98}. It has been proposed that the required flux can arise
from the radiative decays of relic neutrinos of mass $\sim30$~eV which
would also provide the critical density in dark matter
\cite{rs81,dws82}. This hypothesis can naturally account for the
temperature of Lyman-$\alpha$ clouds \cite{dws91} as well as explain
an anomaly found in the abundance of He~I in three high-redshift
Lyman-limit systems of the QSO HS~1700+6416 \cite{dws94}. Moreover,
the decays of the neutrinos in galactic halos such as ours can account
for the observed ionization of the interstellar medium, which again is
difficult to account for by conventional sources \cite{dws93}. Various
astrophysical and cosmological constraints can then pin down the
parameters of the decaying hot dark matter (HDM) cosmology rather
precisely \cite{dws97a}:
\begin{eqnarray}
\label{param}
 m_\nu &=& 27.4 \pm 0.2 \ev \quad \Longrightarrow \quad
 \Omega_\nu h^2 = 0.293 \pm 0.003 , \\ 
 \tau_\nu &\simeq& (1-2) \times 10^{23} {\rm sec} , \\ \nonumber
 h &=& 0.548 \pm 0.003 , \quad {\rm for} \quad \Omega_\nu + \Omega_\B = 1 .
\end{eqnarray}
(Here $\Omega_\nu$ and $\Omega_\B$ denote the fraction of the critical
density in neutrinos and baryons, respectively, and $h$ is the Hubble
constant in units of 100~km~sec$^{-1}$~Mpc$^{-1}$. We note that most
dynamical estimates of the clustered mass density upto supercluster
scales \cite{pjep93} as well as recent observations of Type Ia
supernovae at high redshift \cite{perl98,garn98} actually suggest
$\Omega\sim0.3-0.4$ (with a possible cosmological constant). The
decaying neutrino cosmology can, in principle, accomodate this as long
the constraint (\ref{param}) is respected, i.e. for a high value of
$h$.) Direct searches for the expected UV line at
$E_\gamma=13.7\pm0.1$~eV (i.e. $\lambda=905\pm7~^{0}\!$A) have so far
proved unsuccessful but new results are expected soon from the EURD
detector on the Spanish MINISAT 01 satellite \cite{dws97b}.

As is well known, large-scale structure formation is problematic in a
HDM cosmology because the primordial density perturbation is severely
damped on small scales due to neutrino free-streaming
\cite{bes80}. This would lead to supercluster-size objects forming
first and smaller objects forming subsequently through their
fragmentation, which is in conflict with a variety of observations
\cite{pjep93}. However the problem may be solved if there is an
additional source of small-scale fluctuations, e.g. from a network of
cosmic topological defects \cite{vsb91,gssv93}.

It had been noted some time ago \cite{be84} that reionization of the
IGM, e.g. by early star formation, leads to suppression of the
anisotropy in the cosmic microwave background (CMB) on angular scales
smaller than the horizon size at the reionization epoch. This would
then appear to provide a potential test of the decaying HDM
cosmology. Although much work has been done on reionization in cold
dark matter (CDM) and (isocurvature) baryonic dark matter (BDM) models
\cite{ssv93,hs94,dj95,db97}, these results are not directly applicable
to the present case where the universe is {\em gradually} reionized by
the decaying neutrinos. (An exception is the work of Dodelson \& Jubas
(1995) who considered a variety of possible ionization histories and
estimated the CMB fluctuation signal expected in specific
experiments.) Hence it is necessary to make a specific study of the
problem. Such a calculation has been performed already by Tuluie,
Matzner and Anninos (1995) (see also Anninos et al. 1991) who
numerically simulated the growth of structure in a HDM universe and
followed photon trajectories in the reionized universe in order to
construct CMB temperature maps. However these authors fixed the
amplitude of the initial density perturbation by requiring that the
autocorrelation function of the dark matter particles should equal the
observed value for clusters ($\zeta(r)=(r/r_0)^{-1.8}$) at a redshift
of $z=0$ for separations $r\sim25h^{-1}$~Mpc. Since the COBE discovery
of super-horizon scale CMB fluctuations this is no longer appropriate
and the primordial density perturbation should be normalized via the
COBE quadrupole. It is then seen that the HDM model has too little
power on cluster and smaller scales to match observations and, as
mentioned earlier, some other source of small-scale power must be
sought. Since this issue is model-dependent, the effects of
reionization on the CMB anisotropy ought to be computed without
specific reference to the manner in which structure was generated.

In this Letter we first make the necessary connection between the
ionization history of the IGM and the lifetime of the decaying
neutrinos. We then incorporate this in a numerical code which computes
the CMB anisotropy with high precision. To compare with observational
data, we normalize the primordial density perturbation to COBE and
present the results as an angular power spectrum.

\section{Reionization in the decaying neutrino cosmology}

In the standard cosmology the universe (re)combines at a redshift of
$z_\rec\sim1300$ \cite{pjep}. In particular the last scattering
surface (LSS) of the relic blackbody photons is well approximated by a
gaussian of width $\Delta\,z=78$ located at a redshift $z\sim1065$
\cite{jw85}. The presence of decaying neutrinos generates UV photons
which reionize the IGM, thus significantly broadening the LSS
\cite{srs91}. We study the ionization history from early times by
direct integration of the rate equation for the ionization fraction of
hydrogen, $x_{\H}\equiv\,n_{e}/n_{\H}$, following previous work
\cite{sw84,ags88,dj92}.

At high temperatures both helium and hydrogen are fully ionized and
$x_{\H}\approx1$. Helium (re)combines while hydrogen is still fully
ionized and this process can be approximated by the Saha ionization
equilibrium equation, quite independently of the subsequent evolution.
The evolution of the free electron fraction through hydrogen
(re)combination and reionization is driven by two distinct processes
--- transitions via excited states, and transitions directly to the
ground state,
\begin{eqnarray}
\label{xhevol1}
 -\frac{\df x_\H}{\df t} &=& a \left(\alpha_{1s}\frac{n_e^2}{n_\H} 
 - \sigma_\I \frac{n_{1s} n_\I}{n_\H}\right)_{ep \to \H_{1s}} \\ \nonumber 
 & & \mbox{}+ a C \left(\alpha\frac{n_e^2}{n_\H} 
 - \beta\frac{n_{1s}}{n_\H} \ex^{-B_1/kT_B}\right)_{ep \to \H^{*} \to \H_{1s}},
\end{eqnarray}
where $\alpha_{1s}$ and $\alpha$ are the recombination rates to the
$1s$ level and excited states respectively, $\beta$ is the ionization
rate from the ground state, $C$ is a correction factor \cite{pjep} and
$B_1=13.6$~eV is the ground state binding energy. In the standard
picture of (re)combination of hydrogen, the first component is
negligible because every combining electron--proton pair releases a
photon which immediately ionizes another neutral hydrogen
atom. However it cannot be neglected when there is an an additional
injection of ionizing photons from the decaying neutrinos at a rate
\begin{equation}
 \left(\frac{\df n_\I}{\df t}\right)_{\rm decay} =
\frac{n_\nu}{\tau_\nu} ,
\end{equation}
where $n_\I$ and $n_{\nu}$ are the number density of ionizing photons
and neutrinos respectively and $\tau_\nu$ is the neutrino
lifetime. Then, including the effect of dilution due to expansion and
the redshifting below the Rydberg energy, $n_\I$ evolves in time
according to
\begin{equation}
 \frac{\df n_\I}{\df t} + \left[\sigma_\I n_{1s} 
    + \left(3 + \frac{B_1}{kT}\right) H\right] n_\I 
  = \alpha_{1s} n_e^2 + \frac{n_\nu}{\tau_\nu} ,
\end{equation}
so the equilibrium value of $n_\I$ is  
\begin{equation} 
 (n_\I)_{\rm eq} = \frac{\alpha_{1s}n_e^2 + n_\nu\tau_\nu}
                   {\left[\sigma_\I n_{1s} + (3 + B_1/kT)H\right]} .
\end{equation}
Except for a small region when $x_\H$ is close to unity, $n_\I$ is
well approximated by its equilibrium value \cite{dj92}. It is also a
good approximation to replace the number density $n_{1s}$ of hydrogen
atoms in the $1s$ state by $(1-x_\H)n_\H$. Equation~(\ref{xhevol1})
then reads
\begin{eqnarray}
\label{xhevol2}
\lefteqn{ -\frac{\df x_\H}{\df t} = } \\ \nonumber
 & & a\left[\alpha_{1s} x_\H^2 n_\H 
  - \sigma_\I (1 - x_\H) \left({\alpha_{1s} x_\H^2 n_\H^2 + n_\nu /\tau_\nu} 
  \over {\sigma_\I (1 - x_\H) n_\H + (3 + B_1/kT)H} \right)\right] \\ \nonumber
 & & \mbox{} 
 + a C \left(\alpha x_\H^2 n_\H - \beta (1 - x_\H) \ex^{-B_1/kT_B}\right).
\end{eqnarray}
In addition we must follow the evolution of the matter and radiation
temperatures. We modify the standard evolution as incorporated in the
computer code COSMICS \cite{mb95} by including the energy,
$E_\nu-B_1$, released by the decaying neutrinos (as long as there is
neutral hydrogen present to capture the photons).

\begin{figure}
\centering
\leavevmode\epsfysize=5.4cm \epsfbox{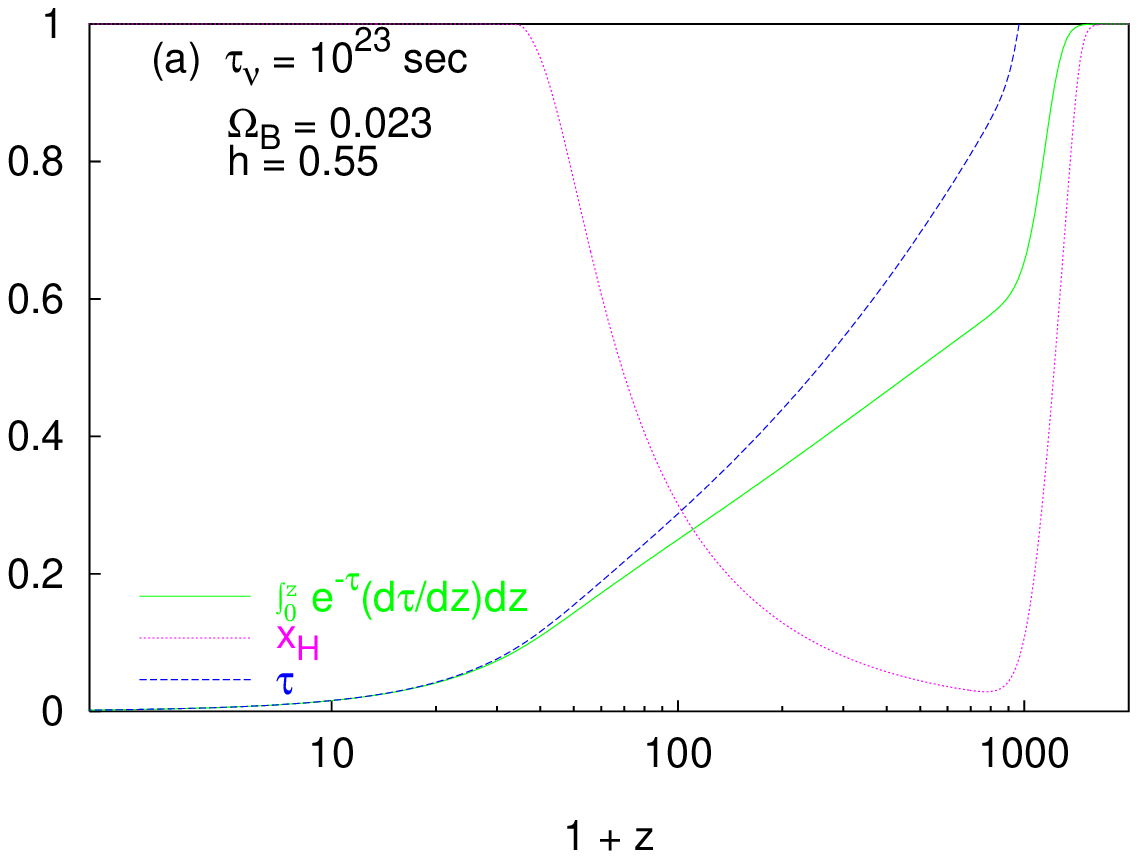}\\
\leavevmode\epsfysize=5.4cm \epsfbox{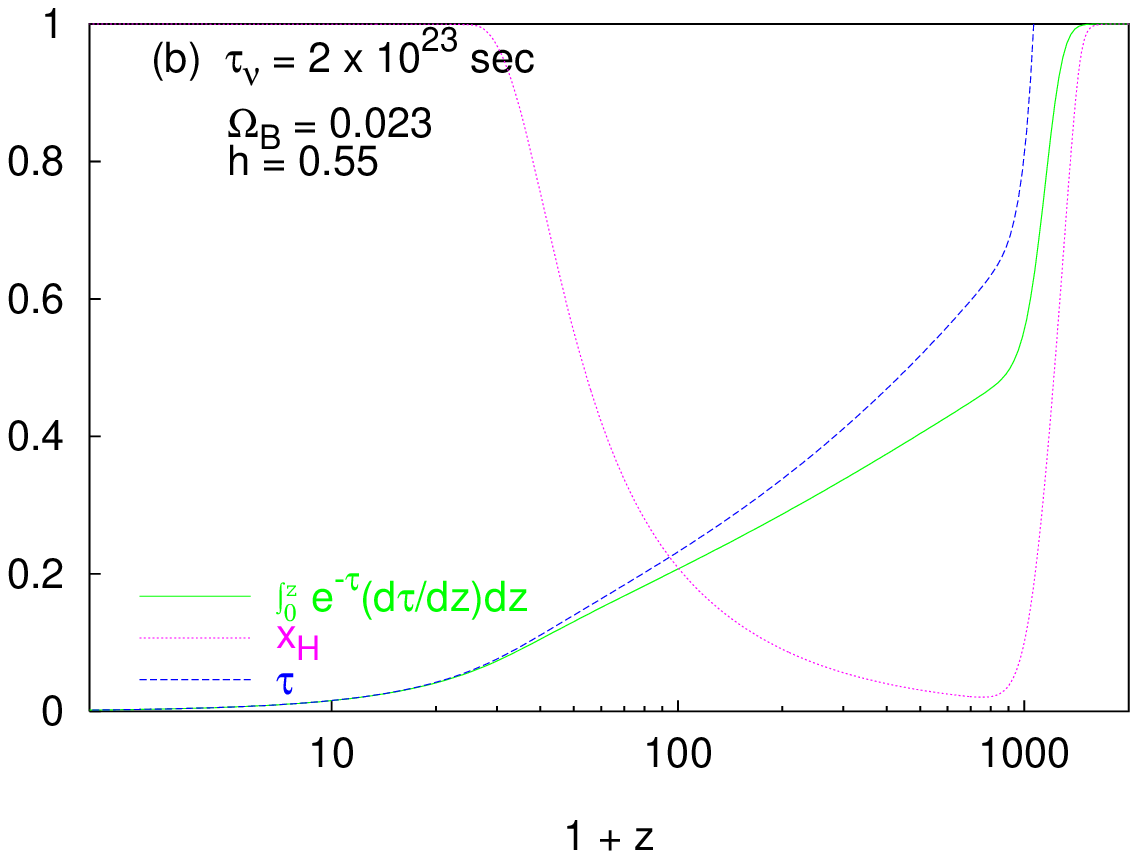}\\
\caption{The fractional ionization, optical depth and cumulative
         visibility function in the decaying neutrino cosmology for a
         neutrino lifetime of (a) $10^{23}$~sec and (b) $2\times10^{23}$~sec.}
\label{fig1}
\end{figure}

We integrate Equation~(\ref{xhevol2}) numerically using a
semi-implicit method. The resulting ionization fraction, optical depth
and cumulative visibility function
($=\int_0^z\,\ex^{-\tau}(\df\,t/\df\,z)\,\df\,z$) are shown in
Figure~\ref{fig1}. For the Hubble parameter we use $h=0.55$
\cite{dws97a}. The baryon density obtains from taking
$\Omega_{\B}h^2=7\times10^{-3}$ as indicated by the primordial
abundance of $^4$He inferred from observations of blue compact
galaxies by Olive, Skillman \& Steigman (1997), the high D abundance
measured in quasar absorption systems \cite{swc97,webb97} and the
abundance of $^7$Li in Pop~II stars \cite{bf97}. If however we adopt
the higher $^4$He abundance obtained by Izotov, Thuan \& Lipovetsky
(1998) in conjunction with the lower D abundance of Burles \& Tytler
(1998) then the appropriate value is
$\Omega_{\B}h^2=1.88\times10^{-2}$. Both values for the baryon
density, viz. $\Omega_\B=0.023,~0.062$ yield a good $\chi^2$ for the
fit to the theoretical expectations \cite{us98}. We also consider two
values for the neutrino lifetime, $\tau_\nu=10^{23}$,
$2\times10^{23}$~sec. It is seen that the IGM is fully ionized for a
redshift below $\sim30$ although the optical depth reaches unity only
at a redshift exceeding $\sim1000$. The LSS is significantly broadened
as is evident from the slow growth of the cumulative visibility
function.

It had been argued \cite{zs69} that the universe must have been
neutral during the epoch $300\Omega_{\B}^{-7/9}<z<z_{\rm rec}$ in
order not to induce a ``y'' distortion in the blackbody spectrum of
the CMB. However this assumed that all matter is constituted of
baryons. Allowing for non-baryonic dark matter, this argument is
evaded and the universe is not required to have (re)combined
\cite{bs91}. We have checked that there is no conflict in the decaying
HDM cosmology with even the very restrictive bound
$y<2.5\times10^{-5}$ from the COBE FIRAS data \cite{cobespec96}.

\section{Calculation of the CMB anisotropy}

As mentioned earlier, the anisotropy in the CMB will be damped as a result
of the broadening of the last scattering surface. To calculate this we
use the CMBFAST code \cite{sz96} incorporating the ionization fraction
evolution described above. The primordial density perturbation is
assumed to be scale-invariant and is normalized on super-horizon
scales to the CMB quadrupole moment measured by COBE: $Q_{\rm
RMS-PS}\simeq18\,\mu$K \cite{cobequad96}.

\begin{figure}
\centering
\leavevmode\epsfysize=5.4cm \epsfbox{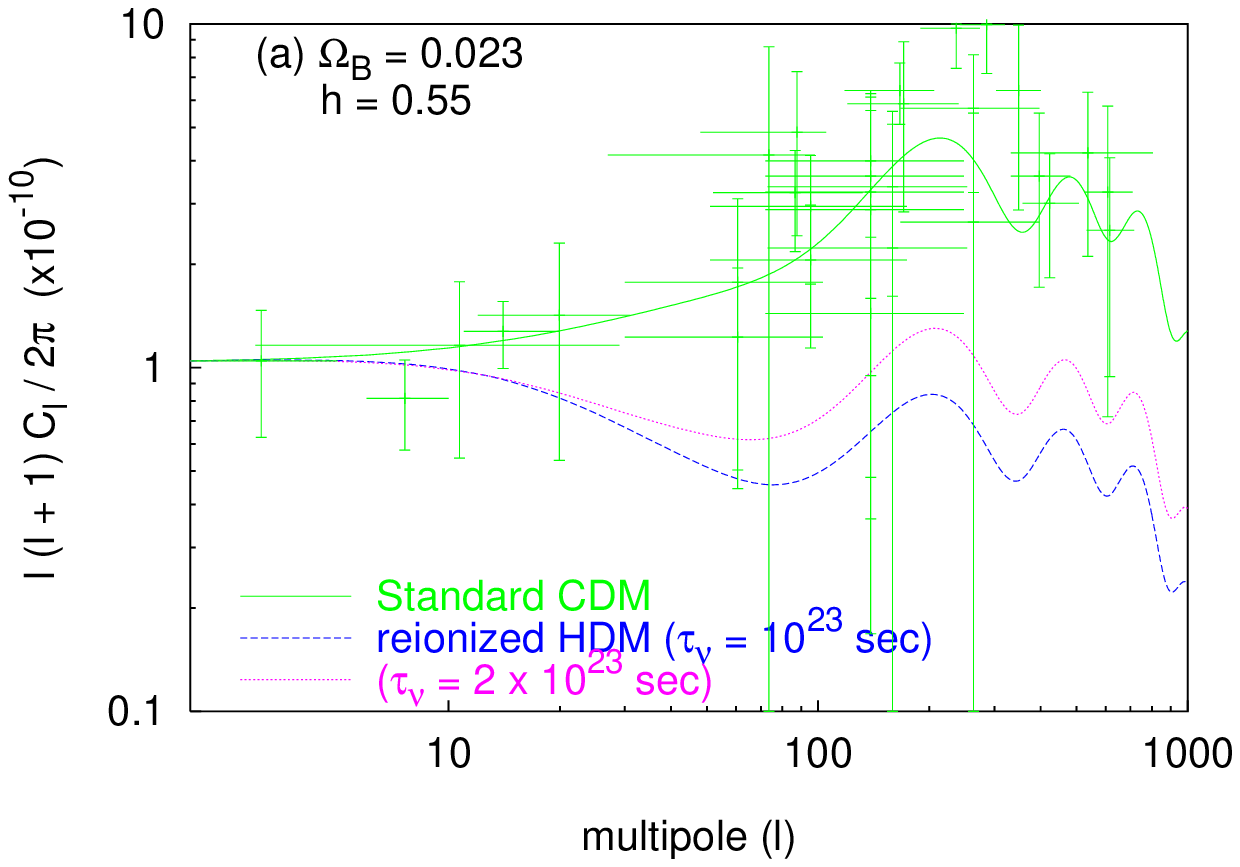}\\
\leavevmode\epsfysize=5.4cm \epsfbox{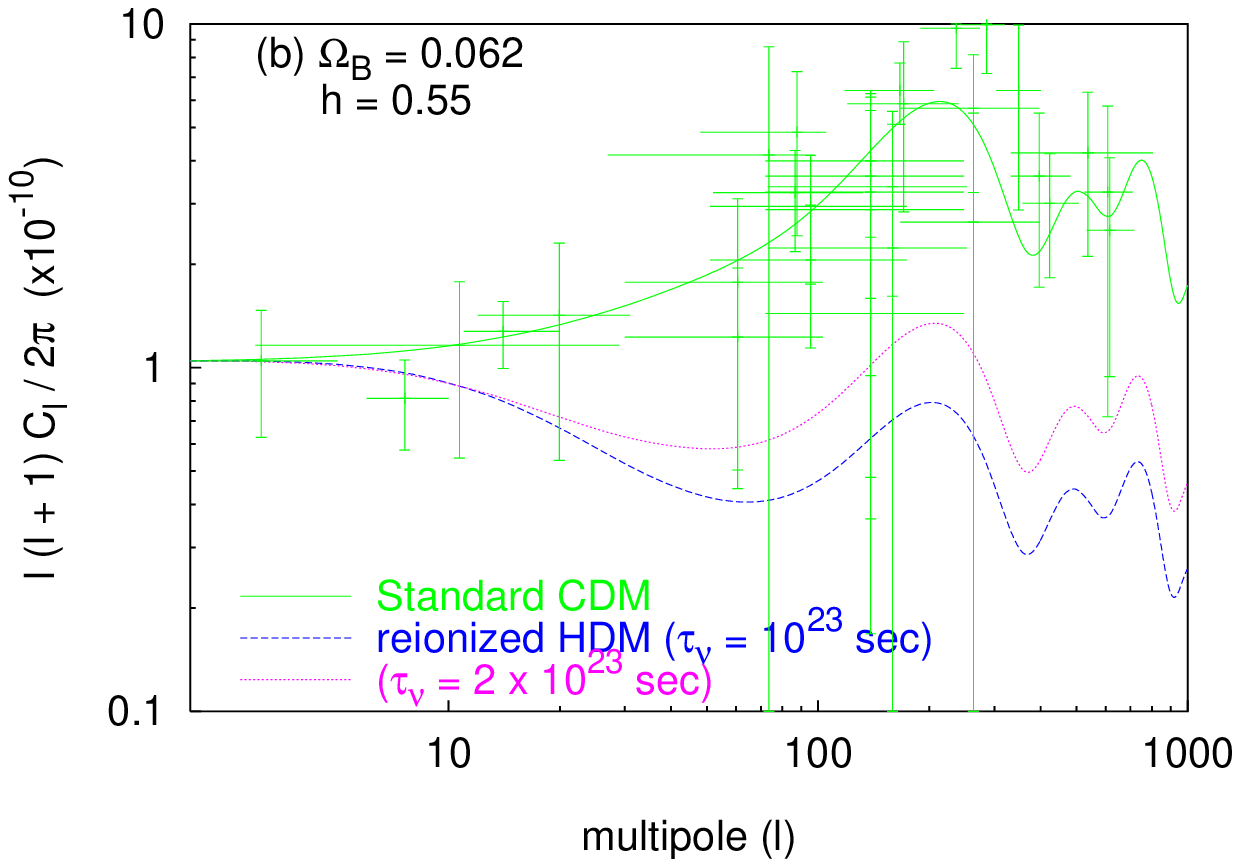}\\
\caption{Angular power spectrum of CMB anisotropy in the decaying
         neutrino cosmology for (a) $\Omega_\B=0.023$ and (b)
         $\Omega_\B=0.062$, with $\tau_\nu=10^{23},\,2\times10^{23}$~sec.}
\label{fig2}
\end{figure}

The calculated angular power spectrum is shown in
Figure~\ref{fig2}. As expected the acoustic peaks are damped by
reionization, with the extent of damping decreasing as the neutrino
lifetime increases. (We stress that the damping effect is largely due,
not to the reionization itself, but rather the presence of more free
electrons at all reshifts which widens the last scattering surface. In
particular, sudden reionization at a late redshift would not result in
significant damping.) An increase in the assumed baryon density alters
the relative level of damping of the second and third acoustic
peaks. A selection of data from recent ground-based and balloon-borne
experiments is also shown for comparison.

To obtain a quantitative measure of the (dis)agreement of the decaying
neutrino prediction with present observations we perform a $\chi^2$
analysis \cite{lbbb97} after convolving the predicted CMB power
spectrum with the experimental window functions. The $\chi^2$ for the
decaying HDM model ranges between $\sim500-700$ for the allowed range
of $\tau_\nu$ and $\Omega_\B$, for 29 degrees of freedom (32 data
points - 2 fitted parameters - 1 COBE normalization). Since there is
less than 0.5\% probability of obtaining a $\chi^2$ exceeding 52.34,
the decaying HDM theory clearly does not fit the trend in the present
data. However it would be premature to draw any strong conclusion from
this, given that there may still be large systematic uncertainties in
the present data. A definitive test will however be possible with the
forthcoming MAP and PLANCK all-sky surveyor satellite missions, which
are expected to reliably measure the small-scale anisotropy with an
accuracy of a few per cent \cite{smoot}. Moreover these experiments
will also measure the CMB polarization, thus providing an independent
constraint on the ionization history \cite{ktps98}.

\section{Conclusions}

We have computed, in a consistent manner, the angular power spectrum
of CMB anisotropy to be expected in the decaying HDM cosmology. The
damping of the acoustic peaks due to the gradual reionization of the
IGM is shown to provide a clear test of the model.

Other possible observational signatures of reionization have been
considered. With regard to the CMB, although large angular-scale
fluctuations are damped, new fluctuations are generated on
arcminute-scales by the motions of the ionized gas --- the Vishniac
effect \cite{ov86,v87}. This has been studied in some detail for both
CDM and BDM models and various constraints inferred from observational
limits on small angular-scale fluctuations
\cite{e88,hss94,dj95,hw96}. For HDM, the small-scale power in the
primordial (inflationary) density perturbation is generically
suppressed which would normally imply a negligible Vishniac
effect. However as we have noted some additional source of small-scale
power such as cosmic topological defects is needed in any case to
allow galaxies to form in an HDM universe. Lacking a precise knowledge
of such fluctuations, it is not possible to make a model-independent
calculation of the Vishniac effect in the present case.

\section*{Acknowledgments}

We thank Charlie Lineweaver for providing the CMB data and
experimental window functions, Ed Bertschinger for the COSMICS code
and Uro\u{s} Seljak and Matias Zaldarriaga for the CMBFAST
code. D.W.S. is grateful to M.U.R.S.T. for their financial support of
this work.

\end{document}